\documentclass[aps,superscriptaddress,twocolumn,floats,prd]{revtex4} 
\usepackage{graphicx} %
\usepackage{epsfig,amsmath}
\usepackage{amssymb}
\usepackage{rotate}
\usepackage{color}
\usepackage{bm}

\DeclareFontFamily{OT1}{pzc}{}
\DeclareFontShape{OT1}{pzc}{m}{it}%
            {<-> s * [1.10] pzcmi7t}{}
\DeclareMathAlphabet{\mathscr}{OT1}{pzc}%
                                {m}{it}

\newcommand\lsim{\mathrel{\rlap{\lower4pt\hbox{\hskip1pt$\sim$}}
        \raise1pt\hbox{$<$}}}
\newcommand\gsim{\mathrel{\rlap{\lower4pt\hbox{\hskip1pt$\sim$}}
        \raise1pt\hbox{$>$}}}

\newcommand{\be}{\begin{equation}}
\newcommand{\ee}{\end{equation}}
\newcommand{\bea}{\begin{eqnarray}}
\newcommand{\eea}{\end{eqnarray}}
\def\ba#1\ea{\begin{align}#1\end{align}}

\newcommand{\refeq}[1]{Eq.~(\ref{eq:#1})}          
\newcommand{\refeqs}[2]{Eqs.~(\ref{eq:#1})--(\ref{eq:#2})}          
          
\newcommand{\reffig}[1]{Fig.~\ref{fig:#1}}

\newcommand{\vs}{\nonumber\\}       

\renewcommand{\v}[1]{\mathbf{#1}}

\newcommand{\vk}{\v{k}}

\newcommand{\<}{\langle}
\renewcommand{\>}{\rangle}

\newcommand{\vnhat}{\v{\hat{n}}}
\newcommand{\vkhat}{\v{\hat{k}}}

\renewcommand{\d}{\delta}
\renewcommand{\a}{\alpha}

\newcommand{\D}{\Delta}

\newcommand{\iMpch}{\,h~{\rm Mpc}^{-1}}

\newcommand{\C}{\mathcal{C}}
\newcommand{\G}{\mathcal{G}}

\renewcommand{\Re}{{\rm Re}} 

\begin{document}

\title{CMB Power Asymmetry from Non-Gaussian Modulation}

\author{Fabian Schmidt}
\affiliation{Theoretical Astrophysics, California Institute of
    Technology, Mail Code 350-17, Pasadena, California  91125}
\affiliation{Department of Astrophysical Sciences, Princeton University, Princeton, NJ 08544, USA}
\affiliation{Einstein Fellow}
\author{Lam Hui}
\affiliation{Department of Physics, Institute for String, Cosmology
  and Astroparticle Physics, Columbia University, New York, NY 10027}

\begin{abstract}
Non-Gaussianity in the inflationary perturbations can couple
observable scales to modes of much longer wavelength (even superhorizon), 
leaving as a signature a large-angle modulation of the observed 
cosmic microwave background (CMB)
power spectrum. This provides an alternative origin
for a power asymmetry which is otherwise often ascribed to a breaking of 
statistical isotropy.  
The non-Gaussian modulation effect can be significant even for typical
$\sim 10^{-5}$ perturbations, while respecting current 
constraints on non-Gaussianity, if the squeezed limit of the
bispectrum is sufficiently infrared divergent.  
Just such a strongly infrared-divergent bispectrum has been claimed
for inflation models with a non-Bunch-Davies initial state, for instance.  
Upper limits on the observed CMB power asymmetry place stringent constraints
on the duration of inflation in such models.
\end{abstract}

\maketitle


Large-scale features in the cosmic microwave background (CMB) offer
interesting avenues for testing phenomena that occurred at very
early times in the Universe's history.  
While most inflationary models predict approximately scale-invariant,
Gaussian fluctuations, some amount of non-Gaussianity is invariably
generated \cite{Maldacena03}.  
In this Letter, we show that even for an almost scale-independent power
spectrum of curvature perturbations (i.e. $\sim 10^{-5}$ in amplitude on all
scales), primordial
non-Gaussianity can lead to interesting, significant effects
on the CMB, in particular a large angular scale modulation of the small-scale
power spectrum. This is achieved without
violating stringent observational bounds on non-Gaussianity
in the subhorizon perturbations.

There are some observational indications for a dipolar
modulation of the CMB power spectrum \cite{HoftuftEtal09,PaciEtal07,EriksenEtal}.  
Such an anisotropic CMB sky can be described by \cite{GordonEtal05}
\ba
\hat \Theta (\vnhat) =\:& [1 + f(\vnhat)] \Theta(\vnhat),
\label{eq:Taniso}
\ea
where $\hat \Theta (\vnhat) $ is the observed, anisotropic 
temperature fluctuation $\delta T/T$, while
$\Theta(\vnhat)$ is a statistically isotropic temperature field, and $f(\vnhat)$ is
the modulating function.  Note that
while $\Theta(\vnhat)$ is statistically isotropic and is thus (statistically)
invariant under a rotation of the coordinate system, $f(\vnhat)$ depends 
on fixed directions on the sky.  

The lowest order modulation is a dipole, as any monopole of $f(\vnhat)$ is 
absorbed in the angle-averaged CMB power spectrum.  The most
recent analysis of \cite{HoftuftEtal09} obtains a statistically significant
dipolar asymmetry, while the WMAP team does not confirm this finding 
\cite{BennettEtal11}.   \citet{HansonEtal} find that beam asymmetries
provide an explanation for the nonzero quadrupolar asymmetry.  
Several scenarios have been proposed in the literature to explain possible
power asymmetries:  \cite{ContaldiEtal03,DonoghueEtal09} 
considered remnants from the preinflationary
phase, \cite{ErickcekEtal1,ErickcekEtal2,LibanovRubakov} proposed a single
large-scale curvature perturbation, while \cite{AckermanEtal} studied a 
spacelike vector field.  These scenarios either involve
 a change in the inflation field $\Delta\varphi \sim A$ across the present 
horizon, many orders of magnitude larger than expected from the amplitude
of fluctuations, or a breaking of the symmetries of the background.

Alternatively, one can interpret a large-scale modulation of the
CMB temperature fluctuations as due to a non-Gaussian coupling 
between long and short wave modes \cite{francis,lewis2011}.  
The power spectrum of
the Bardeen potential $\phi$ on short scales is modulated by the presence
of long modes if the fluctuations are non-Gaussian.  
We can Taylor expand the power spectrum of short modes ($k$)
in the presence of long modes ($k_\ell \ll k$) \cite{Maldacena03}:
\begin{eqnarray}
\label{eq:PphiNG}
&& P_\phi^{\rm mod} (\vk) = P_\phi(k)\left[1 + \int \frac{d^3 k_\ell}{(2\pi)^3} 
G(\vk,\vk_\ell) \phi(\vk_\ell) \right] \vs
&& G(\vk,\vk_\ell) \equiv \frac{B_\phi (|\vk + \vk_\ell/2|, |-\vk +
  \vk_\ell/2|, |-\vk_\ell|)}{P_\phi (k_\ell) P_\phi(k)} \, ,
\end{eqnarray}
where $P_\phi^{\rm mod} (\vk)$ is the modulated power spectrum and
$B_\phi$ is the bispectrum
\footnote{One can also derive the same by
writing the non-Gaussian field as a convolution
of Gaussian fields. See e.g. \cite{fsmk,shmc}.}.  
$G(\vk,\vk_\ell)$ can
be understood as a scale- and orientation-dependent generalization of 
the dimensionless nonlinearity parameter $f_{\rm NL}$.

The scenario we are considering is not statistically anisotropic
in any fundamental sense;  rather, 
the observed power spectrum $P_\phi^{\rm mod} (\vk)$ 
depends on the direction of $\vk$ because the long modes
in our particular realization of the Universe statistically pick out
certain directions $\vk_\ell$, and non-Gaussianity couples
these long modes to the observable ones.  
Also, this effect doesn't require having a large amplitude
long wave mode $\phi({\bf k}_\ell)$; a large kernel $G$ in the squeezed
limit is sufficient.

Observational bounds on primordial non-Gaussianity 
are rather tight \cite{SmithSenatoreZaldarriaga}, which might
lead one to expect that the proposed effect must be small.
The key point is that current observational constraints come
from modes where both $\vk_\ell$ and $\vk$ are
within our horizon.  This is however not necessary for
\refeq{PphiNG} to apply, allowing even superhorizon modes $\vk_\ell$
which we cannot directly measure to have an impact on
observable modes $\vk$ in the form of an anisotropic modulation.  
Two conditions should be met for this effect to be interesting:
(1) the kernel $G$ should be anisotropic, i.e. a nontrivial
function of $\hat\vk\cdot\hat\vk_\ell$; and (2)
$G$ has to grow in the squeezed limit, i.e. scale like $k/k_\ell$ to
some positive power.  
Existing constraints effectively bound $G$ only for moderate ratios of $k/k_\ell$,
while the superhorizon modulation effect is sensitive to larger
ratios.
We will interpret claims of power asymmetries in the literature
as upper limits and use them to constrain
models with such a strong coupling between short and long modes.


\emph{Power asymmetry:} The fluctuations of a statistically isotropic Gaussian field $\Theta(\vnhat)$
are specified through the spherical harmonic coefficients,
$\< \Theta_{lm} \Theta^*_{l'm'}\> = \d_{ll'} \d_{mm'} C_l$.  
Adopting the notation of \cite{WK}, the $\Theta_{lm}$ are related
to the Bardeen potential perturbations $\phi(\vk)$ via
\be
\Theta_{lm} = 4\pi \int \frac{d^3k}{(2\pi)^3} (-i)^l \phi(\vk) \Delta_l(k) Y^*_{lm}(\vkhat),
\label{eq:Theta}
\ee
where $\Delta_l(k)$ is the photon temperature transfer function.  The
power spectrum of the Gaussian temperature fluctuations is then given
by,
\ba
C(l) &\:= \frac{2}{\pi}\int k^2 dk\: P_\phi(k) |\Delta_l(k)|^2.
\label{eq:Cl}
\ea
On the other hand, the spherical harmonic coefficients of the modulated field $\hat \Theta$
[\refeq{Taniso}] are 
\ba
\hat \Theta_{lm} - \Theta_{lm} = \!\!\!
\sum_{LM,l',m'}\!\!\! \Theta_{l'm'} f_{LM} \int d^2\Omega\: Y^*_{lm} Y_{LM} Y_{l'm'},\nonumber
\ea
where we have expressed $f(\vnhat)$ in terms of its multipole moments
(with respect to a fixed coordinate system).  
The integral over three spherical harmonics can be written in terms of
Wigner 3-$j$ symbols, leading at linear order in $f_{LM}$ to 
\ba
&\< \hat \Theta_{lm} \hat \Theta^*_{l'm'}\> = \d_{ll'} \d_{mm'} C_l 
+ \sum_{LM} f_{LM} \G^{ll'L}_{-mm'M} \left[C_{l'} + C_{l} \right] \vs
& \G^{ll'L}_{-mm'M} = (-1)^m\sqrt{\frac{(2l+1)(2l'+1)(2L+1)}{4\pi}} \label{eq:cov_aniso}\\
&\qquad\qquad\quad\times \left(\begin{array}{ccc}
l & l' & L \\
0 & 0 & 0
\end{array}\right)
\left(\begin{array}{ccc}
l & l' & L \\
-m & m' & M
\end{array}\right). \nonumber
\ea
The 3-$j$ symbols entail that $l+l'+L$ is even, $m'-m+M=0$, and that
$|l-l'| \leq L \leq l+l'$.  
The latter condition is particularly relevant since
we are interested in the case where $L$ is much smaller than $l$, $l'$.    
\refeq{cov_aniso} gives the covariance matrix of $\hat\Theta$ in multipole
space in terms of the (fixed) multipole moments $f_{LM}$ and
the statistics of $\Theta$.  As expected, the covariance is not diagonal,
but it is very close to diagonal for $l,l' \gg L$, i.e.,
it is nonzero only if $|l-l'| \leq L$.  


\emph{Non-Gaussianity:} We assume that there is some general non-Gaussianity described to leading order
by a bispectrum $B_\phi$.  
We are interested in the limit $k_\ell \ll H_0 \lesssim k$, where $H_0$ is
the Hubble scale today.  Following \refeq{PphiNG}, we expect
that the presence of long-wavelength modes together with
the mode coupling induced by non-Gaussianity lead to a
breaking of statistical isotropy through the preferred direction $\vk_\ell$. 
Consequently, we now calculate the covariance of the temperature field 
given \refeq{PphiNG}.  
Multiplying \refeq{Theta}, with $\Theta_{l'm'}$, and
integrating over one of the momenta leads to
\ba
& \<\Theta_{lm}\Theta^*_{l'm'}\> \simeq \d_{ll'} \d_{mm'} C_l \vs
&\quad + (4\pi)^2 \int \frac{d^3k_\ell}{(2\pi)^3} \phi(\vk_\ell)
\int \frac{d^3k}{(2\pi)^3} [\D_l(k) \D^*_{l'}(k)] \vs
&\quad\quad\times Y^*_{lm}(\hat k) Y_{l'm'}(\hat k) G(\vk,\vk_\ell) P_\phi(k),
\ea
where we set $|\vk-\vk_\ell| \simeq k$ in the squeezed-limit
approximation (corrections scale as $k_\ell/k$ and higher).  We
obtain
\ba
& \!\!\<\Theta_{lm}\Theta^*_{l'm'}\> = \d_{ll'} \d_{mm'} C_l \vs
& + \int \frac{k_\ell^2 dk_\ell}{(2\pi)^3} \sum_{LM} \G^{ll'L}_{-mm'M}
\C_{ll'}(k_\ell) \phi_{LM}(k_\ell),
\label{eq:cov_NG}
\ea
where we have defined
\ba
\C_{ll'}(k_\ell) =\:& \frac1\pi \int k^2 dk\, [\D_l(k) \D^*_{l'}(k) + \D^*_l(k) \D_{l'}(k)] \vs
& \qquad \times P_\phi(k) G_{L}(k,k_\ell) \vs
G(\vk,\vk_\ell) =\:& \sum_{LM} G_L(k,k_\ell) Y^*_{LM}(\hat k_\ell) Y_{LM}(\hat k),
\label{eq:GLM}\\
\phi_{LM}(k_\ell) =\:& \int d\Omega_{k_\ell} \: \phi(\vk_\ell) Y^*_{LM}(\hat k_\ell),
\label{eq:phiLM}
\ea
using the fact that the kernel $G$ only depends on the angle
between $\vk$ and $\vk_\ell$.  Comparing with \refeq{Cl}, we see that
$\C_{ll}(k_\ell)$ is equal to the temperature power spectrum
obtained when replacing $P_\phi(k) \rightarrow G_L (k,k_\ell) P_\phi(k)$,
i.e. with a different initial power spectrum of curvature fluctuations.  
Thus, apart from the fact that the non-Gaussian covariance involves $\C_{ll'}$,
instead of $\C_{ll}+\C_{l'l'}$, it is identical in structure to the covariance obtained for
the anisotropic field \refeq{cov_aniso} \footnote{In \refeq{cov_aniso}, we have assumed a Gaussian covariance for
the projected quantity $\Theta$, while in \refeq{cov_NG} we are projecting a 
non-Gaussian field, leading to this minor difference.}.  
The fractional difference between 
$\C_{ll'}$ and $\C_{ll},\C_{l'l'}$ is of order $L/l \ll 1$.  We will thus
approximate $\C_{ll'}$ in \refeq{cov_NG} with $(\C_{ll}+\C_{l'l'})/2$.  

\begin{figure}[t]
\centering
\includegraphics[angle=-90,width=0.45\textwidth]{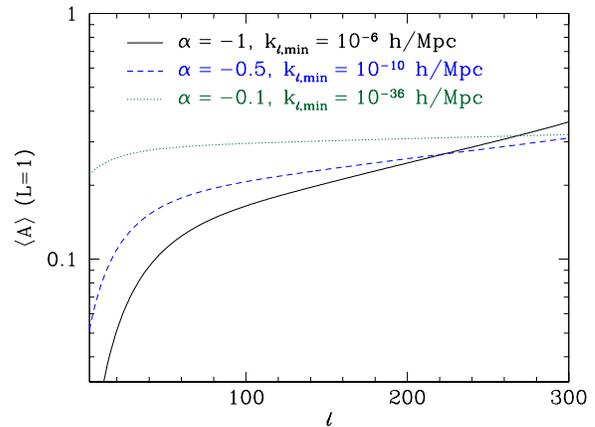}
\caption{Expected amplitude $\<A\>$ of a dipole modulation ($L=1$)
as a function of the CMB multipole $l$, for three 
sets of values for $(\a_1, k_{\ell,\rm min})$ as indicated in the figure.  
We have used \refeq{GLans} with $g_1=1$.  Predictions for different
$L$ can be obtained by multiplying with $g_L(L!)^2 4^L/2(2L)!$.
\label{fig:DeltaCl}}
\end{figure}

We conclude that if $G_L(k,k_\ell)$ is significant in the limit $k_\ell/k \to 0$
for some $L>0$, the temperature fluctuations of the CMB \emph{appear as if they
experience a (large-angle) modulation of multipole order $L$}.  
In particular, this necessitates an anisotropic coupling of
long- and short-wavelength modes.  
We now calculate the amplitude of this modulation.  For scale-free
bispectrum shapes, 
the kernel moments in the squeezed limit ($k_\ell\ll k$) can be written as 
\be
G_L(k,k_\ell) = g_L \left(\frac{k_\ell}{k}\right)^{\alpha_L},
\label{eq:GLans}
\ee
where $g_L$ is a constant and $\alpha_L$ gives the scaling in the
squeezed limit.  We also define 
the temperature power spectrum with a 
tilted spectral index $n_s \rightarrow n_s+\alpha$,
\be
C_l(\alpha) = \frac2\pi \int k^2 dk \left(\frac{k}{k_0}\right)^\alpha\:P_\phi(k) |\D_l(k)|^2,
\ee
where $k_0 = 0.05\,\rm Mpc^{-1}$ is the pivot scale used to 
normalize $P_\phi(k)$.  By comparing \refeq{cov_NG} with \refeq{cov_aniso}, 
we can then read off the relation between the long-wavelength 
perturbations and the anisotropy coefficients $f_{LM}$, for a given $l$ 
considered:
\be
f_{LM} = \frac12 \int\frac{k_\ell^2 dk_\ell}{(2\pi)^3} \phi_{LM}(k_\ell) g_L 
\left(\frac{k_\ell}{k_0}\right)^{\alpha_L} \frac{C_{l}(-\alpha_L)}{C_{l}(0)}.
\label{eq:fLMeff}
\ee
The multipole coefficients which give the amplitude and direction
of the modulation are thus related to the \emph{given realization} of 
the large-wavelength modes
$\phi(\vk_\ell)$.  The last factor in \refeq{fLMeff} encodes the fact that
in general this modulation is $l$-dependent; i.e. one is effectively 
adding a tilted CMB power spectrum $C_l(-\alpha_L)$ 
with angular modulation to the angle-averaged CMB power spectrum.  
While we cannot predict the direction of the power modulation, we
can calculate the expectation value of the \emph{amplitude}, 
defined as $A \equiv (\sum_{M=-L}^L |f_{LM}|^2)^{1/2}$.  
Since the $f_{LM}$ are proportional to $\phi(\vk_\ell)$, they are 
Gaussian-distributed complex numbers with mean zero.  The amplitude $A$ thus follows a 
$\chi$ distribution for $2L+1$ degrees of freedom, with an
expectation value of
\ba
\<A\> =\:& \frac{g_L}{\sqrt{2\pi}}\frac{(L!)^2 4^L}{(2L)!}
\left[\int_{k_{\ell,\rm min}}^{k_{\ell,\rm max}} \!\!\frac{k_\ell^2 dk_\ell}{(2\pi)^3} P_\phi(k_\ell) 
\left(\frac{k_\ell}{k_0}\right)^{2\alpha_L} \right]^{1/2}\vs
&  \times\frac{C_{l}(-\alpha_L)}{C_{l}(0)}
\label{eq:Aexp}
\ea
where we have used
\ba
\< \phi_{LM}(k) \phi^*_{L'M'}(k') \> 
= \d_{LL'} \d_{MM'} (2\pi)^3 \frac{\d_D(k-k')}{k^2}  P_\phi(k).\nonumber
\ea
If $\alpha_L$ is sufficiently negative, $\<A\>$ diverges as we let the lower 
integration bound go to zero.  Such a prediction for $A$ can be ruled out 
to high significance 
by the data if the observational limit $A_{\rm lim} \ll \<A\>$.  
In general, if $P_\phi(k_\ell) \propto k_\ell^{n_s-4}$, then
$\alpha_L < (1-n_s)/2$ for some $L>0$ in \refeq{GLans} is necessary
for a significant large-scale asymmetry of the CMB.  
\reffig{DeltaCl} shows quantitative results for the expected asymmetry 
$\<A\>$ with $L=1$, as a function of the CMB multipole $l$.  
We adopt the ansatz \refeq{GLans} with $g_1=1$, and integrate from 
$k_{\ell,\rm min}$ to $k_{\ell,\rm max} = 1/\eta_{\rm lss}$, where
$\eta_{\rm lss}$ is the comoving distance to the last scattering surface 
(the latter choice is unimportant numerically).
We choose three different sets of $(\a_1,k_{\ell,\rm min})$ and use 
CAMB \cite{camb} for the computation of $C_l(\alpha)$.  
Clearly, a significant amplitude of power asymmetry can be
achieved with a range of parameters.  The closer $\a$ is to zero, the
smaller $k_{\ell,\rm min}$ needs to be to generate a given amount of asymmetry
(at fixed $g_1$).  On the other hand, a more negative $\a_1$ leads to a stronger
scale dependence: the amplitude of the modulation approximately
scales as $l^{-\a_1}$.  

Inflationary bispectra which consist of 
symmetrized polynomials in the three momenta $k_1,\,k_2,\,k_3$ do
not lead to a power asymmetry since the coupling of modes
is isotropic ($G_L = 0$ for $L > 0$).  However, these simple bispectra are 
often only obtained as separable approximations to the more complicated exact 
bispectra, which may themselves in fact lead to $G_L \neq 0$.  Hence,
it is crucial to consider the full, exact bispectrum when determining whether
a given inflationary model leads to a power asymmetry.  
It is clear however that such a power asymmetry requires a violation
of the standard consistency relation \cite{Maldacena03}, at least on the 
scales of interest, as it contains no anisotropic coupling between long 
and short modes.  
A recent example is solid inflation \cite{solid},
which predicts a quadrupolar coupling between long
and short modes. But since in this model $G$ does not
grow in the squeezed limit, the resulting quadrupolar
modulation of the power spectrum is small.

An example of a model that does produce a large-scale power modulation 
is the ekpyrotic (``case II'') scenario
of \cite{KhouryPiazza}, which generates non-Gaussianities that in the 
squeezed limit lead to $\a_1 = -1-\epsilon$ and $\a_2 = -\epsilon$,
where $\epsilon > 0$ is a red tilt.  Thus, in this model one has
divergent dipole and quadrupole modulations.  Another case
which has attracted recent interest is modifications to the initial
state (non-Bunch-Davies) in single-field slow-roll inflation.  These can 
lead to non-Gaussianity with $\a_1 = -1$ 
\cite{ChenEtal,HolmanTolley,MeerburgEtal,Ganc,AgulloParker,AgulloShandera,GancKomatsu}.  
The squeezed bispectrum in the simple non-Bunch-Davies state considered in 
\cite{AgulloShandera} reads
\ba
B_\phi (|\vk + \vk_\ell/2|, |-\vk + \vk_\ell/2|, |-\vk_\ell |) =
\mathcal{B}\, P_\phi (k_\ell) P_\phi (k) \frac{k}{k_\ell}
\nonumber \\
\vspace*{-0.2cm}
\times {\,\rm Re\,} \left[\tilde f_1 \frac{1 - e^{i(1+\mu)k_\ell/k_*}}{1 + \mu} +
 \tilde f_2 \frac{1 - e^{i(1-\mu)k_\ell/k_*}}{1 -\mu}\right],
\label{eq:BnBD}
\ea
where $\Re (\tilde f_1+ \tilde f_2)/2 \approx N_k$, and $N_k$ is the occupation
number of the momentum state $k$
\footnote{This identification ignores interference terms. See \cite{AgulloShandera}.}
,
$\mu = - \hat \vk \cdot \hat \vk_\ell$, $k_* \sim 1/|\eta_{\rm in}|$ is
related to the conformal time at which the initial state is specified,
and $k_\ell > k_*$ in order for this result to apply.  $\mathcal{B}$
is a dimensionless constant equal to 
$4 \epsilon$ in the case studied
in \cite{AgulloShandera,GancKomatsu}, although it could take larger
values in more general models.  
The kernel $G_L$ scales as $k/k_\ell$ in this model,
with $g_L \propto N_k$ for even $L$ and $\alpha_L = -1$.  
For odd $L$, the modulation
scales as $\tilde f_1-\tilde f_2$ which is suppressed by $k_\ell/k$.  
We use observational upper limits on the (primordial) quadrupole modulation amplitude 
$A \lsim 0.1$ \cite{HansonEtal} to place constraints on $k_{\ell,\rm min} = k_*$.  
Numerical evaluation of \refeqs{Aexp}{BnBD} leads to a 95\% C.L. 
lower limit of~\footnote{While \cite{HansonEtal} constrain the asymmetry up to $l=1000$, we conservatively choose $l=500$ here.  We evaluate $P[A < 0.1] > 0.05$ using the $\chi$ distribution.}
\be
k_* \gsim 2 \times 10^{-5} \iMpch\, N_k\, \mathcal{B} \, ,
\ee
implying no more than $\sim 3$~$e$-folds of inflation beyond
our current horizon for $N_k \mathcal{B} \sim 1$. This complements the bound on
a non-Bunch-Davies initial state from backreaction arguments,
which is sensitive to $N_k$ but not ${\cal B}$.  

\emph{Conclusions:} Large-scale modulations of the CMB temperature 
fluctuations offer
an interesting testing ground for the physics of the very early Universe.  
We have shown that certain types of primordial non-Gaussianity
generically predict large power asymmetries in the CMB.
The requisite non-Gaussianity can be thought of 
as an anisotropic, scale-dependent $f_{\rm NL}$ which grows
in the squeezed limit.  

Upper limits on such a modulation can put stringent
constraints on this class of models, which includes scenarios with a 
non-Bunch-Davies initial state.  One can roughly estimate the 
modulation amplitude from
the dimensionless bispectrum amplitude $G(\vk, \vk_\ell)$ for the
longest \emph{observable} mode $k_\ell \sim H_0^{-1}$ through
\be
\<A\> \sim G\left(k, k_\ell = H_0^{-1}\right)\, 4\times
10^{-5}  \left(\frac{H_0}{k_{\ell,\rm min}}\right)^{-\alpha_L} \, ,
\ee
where $k_{\ell, {\rm min}}$ refers to the longest
superhorizon mode responsible for the modulation,
and $\alpha_L$ controls
how fast $G$ grows in the squeezed limit
(Eq. \ref{eq:GLans}).  
Conversely, observational hints of a power asymmetry provide motivation
to further investigate such models.  A convincing 
detection of a CMB power asymmetry, if interpreted in terms of
this scenario, would open an observational window to scales
much larger than the present horizon ($k_\ell \ll 1/\eta_{\rm lss}$), 
which are otherwise completely inaccessible
to direct observation.  This fact distinguishes this effect from
a modulation of the temperature power spectrum by a horizon-scale mode.  

\vspace{-0.03cm}
We have shown that the power asymmetries are generally scale dependent
and increase toward smaller scales.  Thus, unless one invokes a change
in the shape or amplitude of non-Gaussianity on smaller scales, a 
nondetection of a similar 
power asymmetry in the large-scale structure \cite{Hirata09,PullenHirata} puts further
stringent constraints on models that produce such asymmetries.  Furthermore,
models with bispectra that peak more strongly in the squeezed limit than 
the local model will in fact generate a scale-dependent bias 
in large-scale structure tracers \citep{DalalEtal08,fsmk,shmc}
$\Delta b \propto k^{-n}$ with $n > 2$ \cite{GancKomatsu,AgulloShandera}.  
Observations of the 
large-scale structure will thus be of great importance in 
strengthening constraints on the possible non-Gaussian origins of
a power asymmetry. 

\emph{Acknowledgments:} We thank Raphael Flauger, 
Justin Khoury, Alberto Nicolis, and Albert Stebbins for useful discussions.
F.S. was supported by the Moore Foundation at Caltech, and the NASA
Einstein Fellowship Program at Princeton.  L.H. is supported by the DOE and NASA
under Cooperative Agreements No. DE-FG02-92-ER40699 and No. NNX10AN14G.

\vspace{-0.64cm}
\bibliography{NG}

\begin{thebibliography}{32}
\expandafter\ifx\csname natexlab\endcsname\relax\def\natexlab#1{#1}\fi
\expandafter\ifx\csname bibnamefont\endcsname\relax
  \def\bibnamefont#1{#1}\fi
\expandafter\ifx\csname bibfnamefont\endcsname\relax
  \def\bibfnamefont#1{#1}\fi
\expandafter\ifx\csname citenamefont\endcsname\relax
  \def\citenamefont#1{#1}\fi
\expandafter\ifx\csname url\endcsname\relax
  \def\url#1{\texttt{#1}}\fi
\expandafter\ifx\csname urlprefix\endcsname\relax\def\urlprefix{URL }\fi
\providecommand{\bibinfo}[2]{#2}
\providecommand{\eprint}[2][]{\url{#2}}

\bibitem[{\citenamefont{{Maldacena}}(2003)}]{Maldacena03}
\bibinfo{author}{\bibfnamefont{J.}~\bibnamefont{{Maldacena}}},
  \bibinfo{journal}{JHEP} \textbf{\bibinfo{volume}{5}}, \bibinfo{pages}{13}
  (\bibinfo{year}{2003}).

\bibitem[{\citenamefont{{Hoftuft} et~al.}(2009)}]{HoftuftEtal09}
\bibinfo{author}{\bibfnamefont{J.}~\bibnamefont{{Hoftuft}}}
  \bibnamefont{et~al.}, \bibinfo{journal}{\apj} \textbf{\bibinfo{volume}{699}},
  \bibinfo{pages}{985} (\bibinfo{year}{2009}).

\bibitem[{\citenamefont{{Paci} et~al.}(2010)}]{PaciEtal07}
\bibinfo{author}{\bibfnamefont{F.}~\bibnamefont{{Paci}}} \bibnamefont{et~al.},
  \bibinfo{journal}{\mnras} \textbf{\bibinfo{volume}{407}},
  \bibinfo{pages}{399} (\bibinfo{year}{2010}), \eprint{1002.4745}.

\bibitem[{\citenamefont{{Eriksen} et~al.}(2007)}]{EriksenEtal}
\bibinfo{author}{\bibfnamefont{H.~K.} \bibnamefont{{Eriksen}}}
  \bibnamefont{et~al.}, \bibinfo{journal}{\apjl}
  \textbf{\bibinfo{volume}{660}}, \bibinfo{pages}{L81} (\bibinfo{year}{2007}).

\bibitem[{\citenamefont{{Gordon} et~al.}(2005)\citenamefont{{Gordon}, {Hu},
  {Huterer}, and {Crawford}}}]{GordonEtal05}
\bibinfo{author}{\bibfnamefont{C.}~\bibnamefont{{Gordon}}},
  \bibinfo{author}{\bibfnamefont{W.}~\bibnamefont{{Hu}}},
  \bibinfo{author}{\bibfnamefont{D.}~\bibnamefont{{Huterer}}},
  \bibnamefont{and}
  \bibinfo{author}{\bibfnamefont{T.}~\bibnamefont{{Crawford}}},
  \bibinfo{journal}{\prd} \textbf{\bibinfo{volume}{72}},
  \bibinfo{pages}{103002} (\bibinfo{year}{2005}),
  \eprint{arXiv:astro-ph/0509301}.

\bibitem[{\citenamefont{{Bennett} et~al.}(2011)}]{BennettEtal11}
\bibinfo{author}{\bibfnamefont{C.~L.} \bibnamefont{{Bennett}}}
  \bibnamefont{et~al.}, \bibinfo{journal}{\apjs}
  \textbf{\bibinfo{volume}{192}}, \bibinfo{pages}{17} (\bibinfo{year}{2011}).

\bibitem[{\citenamefont{{Hanson} et~al.}(2010)\citenamefont{{Hanson}, {Lewis},
  and {Challinor}}}]{HansonEtal}
\bibinfo{author}{\bibfnamefont{D.}~\bibnamefont{{Hanson}}},
  \bibinfo{author}{\bibfnamefont{A.}~\bibnamefont{{Lewis}}}, \bibnamefont{and}
  \bibinfo{author}{\bibfnamefont{A.}~\bibnamefont{{Challinor}}},
  \bibinfo{journal}{\prd} \textbf{\bibinfo{volume}{81}}, \bibinfo{eid}{103003}
  (\bibinfo{year}{2010}), \eprint{1003.0198}.

\bibitem[{\citenamefont{{Contaldi} et~al.}(2003)\citenamefont{{Contaldi},
  {Peloso}, {Kofman}, and {Linde}}}]{ContaldiEtal03}
\bibinfo{author}{\bibfnamefont{C.~R.} \bibnamefont{{Contaldi}}},
  \bibinfo{author}{\bibfnamefont{M.}~\bibnamefont{{Peloso}}},
  \bibinfo{author}{\bibfnamefont{L.}~\bibnamefont{{Kofman}}}, \bibnamefont{and}
  \bibinfo{author}{\bibfnamefont{A.}~\bibnamefont{{Linde}}},
  \bibinfo{journal}{\jcap} \textbf{\bibinfo{volume}{7}}, \bibinfo{pages}{2}
  (\bibinfo{year}{2003}), \eprint{arXiv:astro-ph/0303636}.

\bibitem[{\citenamefont{{Donoghue} et~al.}(2009)\citenamefont{{Donoghue},
  {Dutta}, and {Ross}}}]{DonoghueEtal09}
\bibinfo{author}{\bibfnamefont{J.~F.} \bibnamefont{{Donoghue}}},
  \bibinfo{author}{\bibfnamefont{K.}~\bibnamefont{{Dutta}}}, \bibnamefont{and}
  \bibinfo{author}{\bibfnamefont{A.}~\bibnamefont{{Ross}}},
  \bibinfo{journal}{\prd} \textbf{\bibinfo{volume}{80}},
  \bibinfo{pages}{023526} (\bibinfo{year}{2009}),
  \eprint{arXiv:astro-ph/0703455}.

\bibitem[{\citenamefont{{Erickcek}
  et~al.}(2008{\natexlab{a}})\citenamefont{{Erickcek}, {Kamionkowski}, and
  {Carroll}}}]{ErickcekEtal1}
\bibinfo{author}{\bibfnamefont{A.~L.} \bibnamefont{{Erickcek}}},
  \bibinfo{author}{\bibfnamefont{M.}~\bibnamefont{{Kamionkowski}}},
  \bibnamefont{and} \bibinfo{author}{\bibfnamefont{S.~M.}
  \bibnamefont{{Carroll}}}, \bibinfo{journal}{\prd}
  \textbf{\bibinfo{volume}{78}}, \bibinfo{pages}{123520}
  (\bibinfo{year}{2008}{\natexlab{a}}), \eprint{0806.0377}.

\bibitem[{\citenamefont{{Erickcek}
  et~al.}(2008{\natexlab{b}})\citenamefont{{Erickcek}, {Carroll}, and
  {Kamionkowski}}}]{ErickcekEtal2}
\bibinfo{author}{\bibfnamefont{A.~L.} \bibnamefont{{Erickcek}}},
  \bibinfo{author}{\bibfnamefont{S.~M.} \bibnamefont{{Carroll}}},
  \bibnamefont{and}
  \bibinfo{author}{\bibfnamefont{M.}~\bibnamefont{{Kamionkowski}}},
  \bibinfo{journal}{\prd} \textbf{\bibinfo{volume}{78}},
  \bibinfo{pages}{083012} (\bibinfo{year}{2008}{\natexlab{b}}),
  \eprint{0808.1570}.

\bibitem[{\citenamefont{Libanov and Rubakov}(2010)}]{LibanovRubakov}
\bibinfo{author}{\bibfnamefont{M.}~\bibnamefont{Libanov}} \bibnamefont{and}
  \bibinfo{author}{\bibfnamefont{V.}~\bibnamefont{Rubakov}},
  \bibinfo{journal}{JCAP} \textbf{\bibinfo{volume}{1011}}, \bibinfo{pages}{045}
  (\bibinfo{year}{2010}).

\bibitem[{\citenamefont{{Ackerman} et~al.}(2007)\citenamefont{{Ackerman},
  {Carroll}, and {Wise}}}]{AckermanEtal}
\bibinfo{author}{\bibfnamefont{L.}~\bibnamefont{{Ackerman}}},
  \bibinfo{author}{\bibfnamefont{S.~M.} \bibnamefont{{Carroll}}},
  \bibnamefont{and} \bibinfo{author}{\bibfnamefont{M.~B.}
  \bibnamefont{{Wise}}}, \bibinfo{journal}{\prd} \textbf{\bibinfo{volume}{75}},
  \bibinfo{eid}{083502} (\bibinfo{year}{2007}),
  \eprint{arXiv:astro-ph/0701357}.

\bibitem[{\citenamefont{{Prunet} et~al.}(2005)}]{francis}
\bibinfo{author}{\bibfnamefont{S.}~\bibnamefont{{Prunet}}}
  \bibnamefont{et~al.}, \bibinfo{journal}{\prd} \textbf{\bibinfo{volume}{71}},
  \bibinfo{eid}{083508} (\bibinfo{year}{2005}).

\bibitem[{\citenamefont{{Lewis}}(2011)}]{lewis2011}
\bibinfo{author}{\bibfnamefont{A.}~\bibnamefont{{Lewis}}},
  \bibinfo{journal}{\jcap} \textbf{\bibinfo{volume}{10}}, \bibinfo{eid}{026}
  (\bibinfo{year}{2011}), \eprint{1107.5431}.

\bibitem[{\citenamefont{{Smith} et~al.}(2009)\citenamefont{{Smith}, {Senatore},
  and {Zaldarriaga}}}]{SmithSenatoreZaldarriaga}
\bibinfo{author}{\bibfnamefont{K.~M.} \bibnamefont{{Smith}}},
  \bibinfo{author}{\bibfnamefont{L.}~\bibnamefont{{Senatore}}},
  \bibnamefont{and}
  \bibinfo{author}{\bibfnamefont{M.}~\bibnamefont{{Zaldarriaga}}},
  \bibinfo{journal}{\jcap} \textbf{\bibinfo{volume}{9}}, \bibinfo{pages}{6}
  (\bibinfo{year}{2009}), \eprint{0901.2572}.

\bibitem[{\citenamefont{{Wang} and {Kamionkowski}}(2000)}]{WK}
\bibinfo{author}{\bibfnamefont{L.}~\bibnamefont{{Wang}}} \bibnamefont{and}
  \bibinfo{author}{\bibfnamefont{M.}~\bibnamefont{{Kamionkowski}}},
  \bibinfo{journal}{\prd} \textbf{\bibinfo{volume}{61}},
  \bibinfo{pages}{063504} (\bibinfo{year}{2000}),
  \eprint{arXiv:astro-ph/9907431}.

\bibitem[{\citenamefont{Lewis et~al.}(2000)\citenamefont{Lewis, Challinor, and
  Lasenby}}]{camb}
\bibinfo{author}{\bibfnamefont{A.}~\bibnamefont{Lewis}},
  \bibinfo{author}{\bibfnamefont{A.}~\bibnamefont{Challinor}},
  \bibnamefont{and} \bibinfo{author}{\bibfnamefont{A.}~\bibnamefont{Lasenby}},
  \bibinfo{journal}{Astrophys. J.} \textbf{\bibinfo{volume}{538}},
  \bibinfo{pages}{473} (\bibinfo{year}{2000}), \eprint{astro-ph/9911177}.

\bibitem[{\citenamefont{Endlich et~al.}(2012)\citenamefont{Endlich, Nicolis,
  and Wang}}]{solid}
\bibinfo{author}{\bibfnamefont{S.}~\bibnamefont{Endlich}},
  \bibinfo{author}{\bibfnamefont{A.}~\bibnamefont{Nicolis}}, \bibnamefont{and}
  \bibinfo{author}{\bibfnamefont{J.}~\bibnamefont{Wang}}
  (\bibinfo{year}{2012}), \eprint{1210.0569}.

\bibitem[{\citenamefont{{Khoury} and {Piazza}}(2009)}]{KhouryPiazza}
\bibinfo{author}{\bibfnamefont{J.}~\bibnamefont{{Khoury}}} \bibnamefont{and}
  \bibinfo{author}{\bibfnamefont{F.}~\bibnamefont{{Piazza}}},
  \bibinfo{journal}{\jcap} \textbf{\bibinfo{volume}{7}}, \bibinfo{pages}{26}
  (\bibinfo{year}{2009}), \eprint{0811.3633}.

\bibitem[{\citenamefont{{Chen} et~al.}(2007)}]{ChenEtal}
\bibinfo{author}{\bibfnamefont{X.}~\bibnamefont{{Chen}}} \bibnamefont{et~al.},
  \bibinfo{journal}{\jcap} \textbf{\bibinfo{volume}{1}}, \bibinfo{eid}{002}
  (\bibinfo{year}{2007}).

\bibitem[{\citenamefont{{Holman} and {Tolley}}(2008)}]{HolmanTolley}
\bibinfo{author}{\bibfnamefont{R.}~\bibnamefont{{Holman}}} \bibnamefont{and}
  \bibinfo{author}{\bibfnamefont{A.~J.} \bibnamefont{{Tolley}}},
  \bibinfo{journal}{\jcap} \textbf{\bibinfo{volume}{5}}, \bibinfo{pages}{1}
  (\bibinfo{year}{2008}).

\bibitem[{\citenamefont{{Meerburg} et~al.}(2009)\citenamefont{{Meerburg}, {van
  der Schaar}, and {Stefano Corasaniti}}}]{MeerburgEtal}
\bibinfo{author}{\bibfnamefont{P.~D.} \bibnamefont{{Meerburg}}},
  \bibinfo{author}{\bibfnamefont{J.~P.} \bibnamefont{{van der Schaar}}},
  \bibnamefont{and} \bibinfo{author}{\bibfnamefont{P.}~\bibnamefont{{Stefano
  Corasaniti}}}, \bibinfo{journal}{\jcap} \textbf{\bibinfo{volume}{5}},
  \bibinfo{eid}{018} (\bibinfo{year}{2009}).

\bibitem[{\citenamefont{{Ganc}}(2011)}]{Ganc}
\bibinfo{author}{\bibfnamefont{J.}~\bibnamefont{{Ganc}}},
  \bibinfo{journal}{\prd} \textbf{\bibinfo{volume}{84}}, \bibinfo{eid}{063514}
  (\bibinfo{year}{2011}), \eprint{1104.0244}.

\bibitem[{\citenamefont{Agullo and Parker}(2011)}]{AgulloParker}
\bibinfo{author}{\bibfnamefont{I.}~\bibnamefont{Agullo}} \bibnamefont{and}
  \bibinfo{author}{\bibfnamefont{L.}~\bibnamefont{Parker}},
  \bibinfo{journal}{Phys.Rev.} \textbf{\bibinfo{volume}{D83}},
  \bibinfo{pages}{063526} (\bibinfo{year}{2011}).

\bibitem[{\citenamefont{Agullo and Shandera}(2012)}]{AgulloShandera}
\bibinfo{author}{\bibfnamefont{I.}~\bibnamefont{Agullo}} \bibnamefont{and}
  \bibinfo{author}{\bibfnamefont{S.}~\bibnamefont{Shandera}}
  (\bibinfo{year}{2012}), \eprint{1204.4409}.

\bibitem[{\citenamefont{Ganc and Komatsu}(2012)}]{GancKomatsu}
\bibinfo{author}{\bibfnamefont{J.}~\bibnamefont{Ganc}} \bibnamefont{and}
  \bibinfo{author}{\bibfnamefont{E.}~\bibnamefont{Komatsu}}
  (\bibinfo{year}{2012}), \eprint{1204.4241}.

\bibitem[{\citenamefont{{Hirata}}(2009)}]{Hirata09}
\bibinfo{author}{\bibfnamefont{C.~M.} \bibnamefont{{Hirata}}},
  \bibinfo{journal}{\jcap} \textbf{\bibinfo{volume}{9}}, \bibinfo{pages}{11}
  (\bibinfo{year}{2009}), \eprint{0907.0703}.

\bibitem[{\citenamefont{{Pullen} and {Hirata}}(2010)}]{PullenHirata}
\bibinfo{author}{\bibfnamefont{A.~R.} \bibnamefont{{Pullen}}} \bibnamefont{and}
  \bibinfo{author}{\bibfnamefont{C.~M.} \bibnamefont{{Hirata}}},
  \bibinfo{journal}{\jcap} \textbf{\bibinfo{volume}{5}}, \bibinfo{pages}{27}
  (\bibinfo{year}{2010}).

\bibitem[{\citenamefont{{Dalal} et~al.}(2008)\citenamefont{{Dalal}, {Dor{\'e}},
  {Huterer}, and {Shirokov}}}]{DalalEtal08}
\bibinfo{author}{\bibfnamefont{N.}~\bibnamefont{{Dalal}}},
  \bibinfo{author}{\bibfnamefont{O.}~\bibnamefont{{Dor{\'e}}}},
  \bibinfo{author}{\bibfnamefont{D.}~\bibnamefont{{Huterer}}},
  \bibnamefont{and}
  \bibinfo{author}{\bibfnamefont{A.}~\bibnamefont{{Shirokov}}},
  \bibinfo{journal}{\prd} \textbf{\bibinfo{volume}{77}},
  \bibinfo{pages}{123514} (\bibinfo{year}{2008}), \eprint{0710.4560}.

\bibitem[{\citenamefont{{Schmidt} and {Kamionkowski}}(2010)}]{fsmk}
\bibinfo{author}{\bibfnamefont{F.}~\bibnamefont{{Schmidt}}} \bibnamefont{and}
  \bibinfo{author}{\bibfnamefont{M.}~\bibnamefont{{Kamionkowski}}},
  \bibinfo{journal}{\prd} \textbf{\bibinfo{volume}{82}},
  \bibinfo{pages}{103002} (\bibinfo{year}{2010}), \eprint{1008.0638}.

\bibitem[{\citenamefont{Scoccimarro et~al.}(2012)\citenamefont{Scoccimarro,
  Hui, Manera, and Chan}}]{shmc}
\bibinfo{author}{\bibfnamefont{R.}~\bibnamefont{Scoccimarro}},
  \bibinfo{author}{\bibfnamefont{L.}~\bibnamefont{Hui}},
  \bibinfo{author}{\bibfnamefont{M.}~\bibnamefont{Manera}}, \bibnamefont{and}
  \bibinfo{author}{\bibfnamefont{K.~C.} \bibnamefont{Chan}},
  \bibinfo{journal}{Phys.Rev.} \textbf{\bibinfo{volume}{D85}},
  \bibinfo{pages}{083002} (\bibinfo{year}{2012}), \eprint{1108.5512}.

\end{thebibliography}

\end{document}